\documentclass[showpacs,pra,a4paper,floatfix,twocolumn]{revtex4}
\usepackage{graphicx}
\usepackage{bm,epsfig}
\usepackage{dsfont}
\usepackage{amsmath}
\usepackage{amssymb}

\newcommand{\ket}[1]{\ensuremath{| #1\rangle}}
\newcommand{\bra}[1]{\ensuremath{\langle #1 |}}

\begin{document}

\title{Nonlinear Effects in Pulse Propagation through \\
Doppler-Broadened Closed-Loop Atomic Media}

\author{Robert \surname{Fleischhaker}}
\email{robert.fleischhaker@mpi-hd.mpg.de}
\affiliation{Max-Planck-Institut f\"ur Kernphysik, Saupfercheckweg 1, 
D-69117 Heidelberg, Germany}

\author{J\"org \surname{Evers}}
\email{joerg.evers@mpi-hd.mpg.de}
\affiliation{Max-Planck-Institut f\"ur Kernphysik, Saupfercheckweg 1, 
D-69117 Heidelberg, Germany}

\date{\today}

\begin{abstract}
Nonlinear effects in pulse propagation through a medium consisting of  four-level double-$\Lambda$-type systems are studied theoretically. We apply three continous-wave driving fields and a pulsed probe field such that they form a closed interaction loop. Due to the closed loop and the finite frequency width of the probe pulses  the multiphoton resonance condition cannot be fulfilled, such that a time-dependent analysis is required. By identifying the different underlying physical processes we determine the parts of the solution relevant to calculate the linear and nonlinear response of the system. We find that the system can exhibit a strong intensity dependent refractive index with small absorption over a range of several natural linewidths. For a realistic example we include Doppler and pressure broadening and calculate the nonlinear selfphase modulation in a gas cell with Sodium vapor and Argon buffer gas. We find that a selfphase modulation of $\pi$ is achieved after a propagation of few centimeters through the medium while the absorption in the corresponding spectral range is small.
\end{abstract}


%
%
%

\maketitle

\section{\label{intro}Introduction}
A main interest in laser driven atomic media is the study of their coherence properties. Coherence effects like electromagnetically induced transparency (EIT)~\cite{eit}, coherent population trapping~\cite{cpt}, lasing without inversion~\cite{lwi}, and others~\cite{FiSw2005,scullybook} are examples where the optical properties of an atomic medium are influenced with coherent fields. The interference of different excitation channels is the main underlying principle here. A particular class of systems in which quantum mechanical interference plays a major role are the so-called closed-loop systems~\cite{closed-loop,veer,hinze,maichen,korsunsky,merriam,schroedermorigi,kajari,windholz,malinovsky,shpaism,mahmoudi}. In these systems the laser-driven transitions form a closed interaction loop such that photon emission and absorption can take place in a cycle. This leads to interference of indistinguishable transition pathways between different states. One consequence of this is that it can render the system dependent on the relative phase of the driving fields. At the same time, however, the investigation of closed-loop systems is made difficult by the fact that the interfering pathways typically prevent the system from reaching a time-independent steady state. Such a stationary state in general is only reached when the so-called multiphoton resonance condition on the detunings of the different driving field is fulfilled, which was therefore assumed in most previous studies. For general laser field detunings, a time-dependent analysis is mandatory~\cite{maichen,mahmoudi}.

Laser driven atomic media are also known to exhibit significant nonlinear optical properties~\cite{boydbook,bjorkholm,harris-nonlin,hemmer,korsunsky,shpaism,merriam,maichen,hinze,kajari,schroedermorigi,windholz,braje,lukin,nonlin-n,matsko,schmidt,hau,nakajima,gong,dey}.
A particular example is the occurrence of an intensity dependent refractive index, with applications such as beam focussing, pulse compression, selfphase- or cross-phase modulation or optical switching~\cite{bjorkholm,nonlin-n,matsko,schmidt,hau,nakajima,gong,dey}. Here, the connection to coherence properties is the following. While an atomic resonance can greatly enhance nonlinear effects in atomic media, the accompanying linear absorption of the same resonance typically renders the medium opaque to the probe field. This can be overcome by tailoring the response via coherence and interference effects. An advantageous situation arises, e.g., if the linear absorption vanishes due to destructive interference while the nonlinear effect is enhanced by constructive interference.

\begin{figure}[t]
\includegraphics[width=6cm]{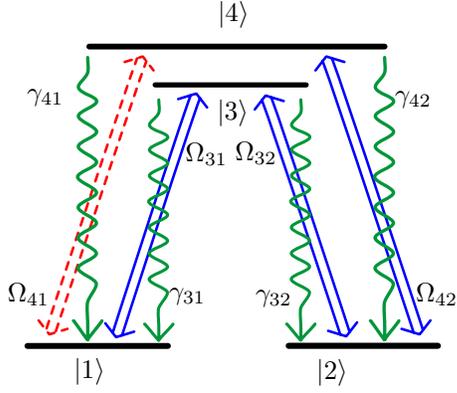}
\caption{\label{fig-system}(Color online)
The four-level atomic system with the four dipole-allowed transitions forming a closed-loop double-$\Lambda$ type scheme. Three transitions are driven by continuous-wave control fields indicated by the solid blue double arrows. The fourth transition couples to the pulsed probe field indicated by the dashed red double arrow. The coupling strengths are given by the Rabi frequencies $\Omega_{jk}$. The spontaneous decays with rates $\gamma_{jk}$ are denoted by the wiggly green lines ($j\in \{3,4\}$, $k\in \{1,2\}$).
}
\end{figure}

Motivated by this, we investigate nonlinear effects in pulse propagation through a closed-loop atomic medium. In particular, we study a four-level atomic system where the four dipole-allowed transitions form a double-$\Lambda$ type scheme (see Fig.~\ref{fig-system}). Three of the fields are assumed to be continous-wave coupling laser fields, while the fourth field is a pulsed probe field. We use a time-dependent analysis, as the multiphoton resonance condition cannot be applied due to the finite frequency spectrum of the probe pulses. The medium is modelled as a dilute gas vapor including Doppler and pressure broadening and an additional buffer gas using realistic parameters. Our main observable is the nonlinear index of refraction of the medium. We find that our system exhibits a high nonlinear index of refraction with small linear and non-linear absorption over a spectral range of several natural linewidths. In this spectral region of interest, the real part of linear and non-linear susceptibility show linear dispersion, such that pulse shape distortions are minimized. For Sodium atoms with Argon buffer gas, we obtain a nonlinear selfphase modulation of $\pi$ after $2.9$~cm of passage through the medium.

The paper is organized as follows. In the following Sec.~\ref{ham} we present our model. In Sec.~\ref{sol}, we solve for the time-dependent long-time limit arising from the closed interaction loop in the form of a series. The interpretation of the series coefficients with respect to their physical meaning (Sec.~\ref{pro}) will enable us to identify the quantities necessary to calculate the linear and nonlinear susceptibility for the probe field of our system (Sec.~\ref{sus}). Doppler and pressure broadening are discussed in Secs.~\ref{dop} and~\ref{buf}. Our results  are presented in Sec.~\ref{res}, both with and without broadening. Finally, Sec.~\ref{con} discusses and summarizes our results.

\section{\label{the}Theoretical Analysis}

\subsection{\label{ham}The model}
In this section we present the Hamiltonian for the four-level system and the interaction with the coupling fields in a suitable interaction picture. We write the field coupling to transition $\ket{j} \leftrightarrow \ket{k}$ ($j\in \{3,4\}$, $k\in \{1,2\}$) as
\begin{align}
 \bm{E}_{jk} = \frac{E_{jk}}{2} \left( \hat{\bm{e}}_{jk} e^{-i \omega_{jk} t} + \textrm{ c.c.}\right) \, ,
\end{align}
with amplitude $E_{jk}$, unit polarization vector $\hat{\bm{e}}_{jk}$, and frequency $\omega_{jk}$. 
For better readability we suppress the space-dependence of the fields. The Hamiltonian in dipole and rotating-wave approximation reads \cite{FiSw2005,scullybook}
\begin{align}
 H = & \sum_{j = 1}^{4} \hbar \omega_j A_{jj} \nonumber \\
 & - \sum_{j = 3}^{4} \sum_{k = 1}^{2} \frac{\hbar \Omega_{jk}}{2} \left\{ e^{-i (\omega_{jk} t - \phi_{jk})} A_{jk} + \textrm{ H.c.} \right\}\,.
\end{align}
The energy of level $|j\rangle$ is denoted by $\hbar \omega_j$ and we have introduced Rabi frequencies $\Omega_{jk}~=~E_{jk} |\hat{\bm{e}}_{jk}~\cdot~\bm{d}_{jk}| / \hbar$ with $\bm{d}_{jk}$ being the dipole matrix element of transition $\ket{j} \leftrightarrow \ket{k}$ ($j\in \{3,4\}$, $k\in \{1,2\}$). The complex phase of the Rabi frequencies was included into the exponential function where $\phi_{jk} = \textrm{arg}(\hat{\bm{e}}_{jk} \cdot \bm{d}_{jk})$. The atomic transition operator is defined as $A_{jk} = \ket{j}\bra{k}$.

The canonical approach with a Hamiltonian of the sort we have just introduced would be to transform it into an interaction picture where the time dependence fully vanishes. Unfortunately, this is not possible in our case. Due to the closed interaction loop, in general a residual time dependence in the Hamiltonian remains. Physically, this means that we cannot expect the system to reach a true stationary state in the long time limit. The best we can do is to use a unitary transformation that gathers all the time dependence in a single exponential factor in front of the probe field Rabi frequency. In this interaction picture we obtain
\begin{align}
\label{hamI}
 H_I = & \hbar (\Delta_{32}-\Delta_{31}) A_{22} - \hbar \Delta_{31} A_{33} \nonumber \\
 & + \hbar (\Delta_{32} - \Delta_{31} - \Delta_{42}) A_{44} \nonumber \\
 & - \frac{\hbar}{2} \left (\Omega_{31} A_{31} + \Omega_{32} A_{32} + \Omega_{42} A_{42} \right. \nonumber \\
 & \left. + \Omega_{41} A_{41} e^{-i (\Delta t - \phi)} + \textrm{ H.c.} \right)\,,
\end{align}
where the detunings are defined as $\Delta_{jk} = \omega_{jk} - (\omega_j - \omega_k)$. We have also defined the so-called multiphoton detuning and an equivalent combination of the dipole phases
\begin{subequations}
\begin{align}
 \Delta = & \Delta_{41} + \Delta_{32} - \Delta_{31} - \Delta_{42}\,, \\
 \phi = & \phi_{41} + \phi_{32} - \phi_{31} - \phi_{42}\,.
\end{align}
\end{subequations}
The multiphoton detuning is a typical quantity characterizing a system with a closed interaction loop. Its significance will become more apparent in Sec. \ref{pro}.

We now set up the master equation for the atomic density matrix $\varrho$. We include the unitary evolution due to the  Hamiltonian in the interaction picture and relaxation dynamics due to spontaneous decay in Born-Markov approximation. The collision induced dynamics will be considered in Sec. \ref{buf}. The unitary evolution is given by the Von-Neumann equation and the spontaneous decay can be written in Lindblad form \cite{FiSw2005}. The master equation in the interaction picture then reads
\begin{align}
\label{meqn}
 \partial_t \varrho^I = & \frac{1}{i \hbar} \left[ H_I, \varrho^I \right] \nonumber \\
 & - \sum_{j = 3}^{4} \sum_{k = 1}^{2} \frac{\gamma_{jk}}{2} \left\{ \left[ \varrho^I A_{jk}, A_{kj} \right] + \textrm{ H.c.} \right\}\,,
\end{align}
where $\varrho^I$ is the density matrix in the interaction picture and $\gamma_{jk}$ is the radiative decay rate of transition $\ket{j} \leftrightarrow \ket{k}$.
For the further analysis we rewrite the master equation in a matrix-vector form. Because the trace of the density matrix is conserved we use the corresponding condition
\begin{align}
 \sum_{j = 1}^{4} \varrho_{jj}^I = 1
\end{align}
to eliminate the diagonal element $\varrho_{44}$. Here, $\varrho^I_{jk} = \bra{j} \varrho^I \ket{k}$. Introducing the vector $R = (\varrho_{11}^I, \varrho_{12}^I, \varrho_{13}^I, \ldots, \varrho_{43}^I)^T$ containing the remaining fifteen elements of the density matrix we find
\begin{align}
 \label{eom}
 \partial_t R + \Sigma = M R\,,
\end{align}
with an inhomogeneous part $\Sigma$ that stems from the elimination of $\varrho_{44}$ and a coefficient matrix $M$. Both $\Sigma$ and $M$ can be directly derived from the master Eq.~(\ref{meqn}) and contain the explicit time dependence arising from the time dependent Hamiltonian Eq.~(\ref{hamI}). The explicit form of $M$ and $\Sigma$ is given in the appendix.

\subsection{\label{sol}Time-Dependent Solution}
To treat the explicit time dependence of the equation of motion we first separate $\Sigma$ and $M$ into the time independent part and the explicitly time dependent part. For this, we define
\begin{subequations}
\label{sm-decomp}
\begin{align}
 \label{sep}
 \Sigma = & \Sigma_0 + \Sigma_{-1} \Omega_{41} e^{i (\Delta t - \phi)} + \Sigma_1 \Omega_{41} e^{-i (\Delta t - \phi)}\,,\\
 M = & M_0 + M_{-1} \Omega_{41} e^{i (\Delta t - \phi)} + M_1 \Omega_{41} e^{-i (\Delta t - \phi)}\,,
\end{align}
\end{subequations}
with time-independent $\Sigma_{j}$ and $M_{j}$ ($j\in\{0,\pm 1\}$).
We see that under the condition $\Delta = 0$ the explicit time dependence vanishes. This is the so-called multiphoton resonance condition. For fixed coupling field frequencies this condition can only be fulfilled for a single probe field detuning $\Delta_{41}$. But we want to investigate probe fields consisting of pulses with finite temporal length, which due to the Fourier relations implies that a whole spectrum of probe field frequencies interacts with the medium at the same time. Thus, we cannot assume the multiphoton resonance condition to be fulfilled~\cite{mahmoudi}. Instead, we have to solve Eq.~(\ref{eom}) including the explicit time dependence. To do so, we expand $R$ as a power series in $\Omega_{41}$,
\begin{align}
\label{r-exp}
 R = & \sum_{n = 0}^{\infty} R_n \Omega_{41}^n \,.
\end{align}
If we assume that the probe field strength is small compared to the control fields this series will converge. Inserting Eqs.~(\ref{sm-decomp}) and (\ref{r-exp}) in  Eq.~(\ref{eom}), we can derive equations of motion for the individual coefficients $R_n$. In order $\mathcal{O}[\Omega_{41}^n]$ we find
\begin{align}
 \partial_t R_n = & M_0 R_n \nonumber \\
 & + \delta_{n,1} \left( \Sigma_{-1} e^{i (\Delta t - \phi)} + \Sigma_1 e^{-i (\Delta t - \phi)} \right) \nonumber \\
 & + \left( M_{-1} e^{i (\Delta t - \phi)} + M_1 e^{-i (\Delta t - \phi)} \right) R_{n-1}\,.
\end{align}
This is an equation for $R_n$ where the coefficient matrix $M_0$ is time independent and only the inhomogeneous part is time dependent. This time dependence is twofold, first again explicitly because of the exponential functions and second because of the dependence on $R_{n-1}$. Thus, we make an ansatz for the solution and write $R_n$ in a Fourier series,
\begin{align}
 R_n = & \sum_{m = -\infty}^{\infty} R_n^{(m)} e^{-i m (\Delta t - \phi)} \,.
\end{align}
Projecting on the Fourier basis functions we derive a hierarchy of time independent equations for the coefficients $R_n^{(m)}$.
Up to order $\mathcal{O}[\Omega_{41}^3]$ we find
\begin{subequations}
 \label{hie}
\begin{align}
  R_0^{(0)} = & M_0^{-1} \Sigma_0 \, ,\\
  R_1^{(\pm 1)} = & \left ( M_0 \pm i \Delta {\mathds 1}\right )^{-1}\left ( \Sigma_{\pm 1} - M_{\pm 1} R_0^{(0)}\right ) \, ,\\
  R_2^{(0)} = & -M_0^{-1}\left ( M_{-1} R_1^{(1)} + M_1 R_1^{(-1)} \right) \, ,\\
  R_2^{(\pm 2)} = & - \left ( M_0 \pm 2 i \Delta {\mathds 1} \right)^{-1} M_{\pm 1} R_1^{(\pm 1)} \, ,\\
  R_3^{(\pm 1)} = & -\left ( M_0 \pm i \Delta {\mathds 1} \right)^{-1} \nonumber \\
 & \times \left ( M_{\pm 1} R_2^{(0)} + M_{\mp 1} R_2^{(\pm 2)}\right ) \, ,\\
  R_3^{(\pm 3)} = & -\left ( M_0 \pm 3 i \Delta {\mathds 1} \right)^{-1} M_{\pm 1} R_2^{(\pm 2)} \, ,
 \end{align}
\end{subequations}
where $\mathds 1$ is the unit matrix and all  other $R_n^{(m)}$ up to this order vanish. In general we find that
\begin{align}
  R = & \sum_{n = 0}^{\infty} \sum_{\genfrac{}{}{0pt}{}{m = -n,}{-n + 2, \ldots}}^{n} R_n^{(m)}\: \Omega_{41}^n\: e^{-i m (\Delta t - \phi)}\,.
\end{align}
Since Fourier coefficients $R_{n}^{(m)}$ in Eq.~(\ref{hie}) only depend on Fourier coefficients $R_{n-1}^{(m)}$ of the next lower order, the full solution can be calculated recursively.

\subsection{\label{pro}Physical Interpretation}
To physically interpret the meaning of the different coefficients we study the influence of the different parts of the solution on the probe field. First, we write down the expansion series for the relevant probe field coherence in the Schr\"odinger picture $\varrho_{41}$ using the explicit transformation relation connecting the Schr\"odinger picture with our interaction picture. We find 
\begin{align}
 \varrho_{41} = & \varrho_{41}^I \:e^{-i (\omega_{41} t - \phi_{41})}\: e^{i (\Delta t - \phi)}.
\end{align}
With $\varrho_{41}^I$ given as component of the solution for $R$ we find
\begin{align}
\label{rho41}
 \varrho_{41} = & \sum_{n = 0}^{\infty} \sum_{\genfrac{}{}{0pt}{}{m = -n,}{-n + 2, \ldots}}^{n} \left[R_n^{(m)}\right]_{13} \:\Omega_{41}^n \nonumber \\
 & \times e^{-i \left[\omega_{41} + (m- 1) \Delta \right] t}\: e^{i \left[\phi_{41} + (m - 1)\phi \right]}\,,
\end{align}
where $[R_n^{(m)}]_{13}$ refers to the thirteenth component of vector $R_n^{(m)}$. Thus, coefficient $[R_n^{(m)}]_{13}$ gives a contribution at the probe field frequency $\omega_{41}$ plus a frequency shift of $(m-1) \Delta$. The corresponding physical process can be identified as follows. 
A combination of dipole phases $\phi = \phi_{41} - \phi_{42} + \phi_{32} - \phi_{31}$ indicates a full evolution through a loop which extends from state $\ket{1}$ to $\ket{4}$ and via $\ket{2}$ and $\ket{3}$ back to state $\ket{1}$. The transition direction is given by the sign of the corresponding dipole phase. The evolution around the interaction loop is also the physical reason for the frequency shift $\Delta$ of such a process. Altogether, $[R_n^{(m)}]_{13}$ represents a process with $m - 1$ loop cycles where the sign of $m - 1$ defines the direction, clockwise for positive or counter-clockwise for negative sign. The remaining $n - (m - 1)$ probe transitions can be interpreted as direct transitions.

\subsection{\label{sus}Linear and Non-Linear Susceptibility}
With the above interpretation we can easily identify the parts of the solution leading to the linear and nonlinear susceptibility in the probe field. Because both contributions should oscillate at the probe field frequency we see that $m = 1$ must be fulfilled in Eq.~(\ref{rho41}). The order of $\Omega_{41}$ enables one to identify
\begin{subequations}
\begin{align}
 \chi^{(1)}(\omega_{41}) \propto & \left[R_1^{(1)} \right]_{13} \quad \text{at} \quad \mathcal{O}\left[\Omega_{41}^1 \right]\,,\\
 \chi^{(3)}(\omega_{41}) \propto & \left[R_3^{(1)} \right]_{13} \quad \text{at} \quad \mathcal{O}\left[\Omega_{41}^3 \right]\,.
\end{align}
\end{subequations}
There is no second order contribution to the susceptibility as it should be for an isotropic medium~\cite{boydbook}. By comparing the microscopically calculated value for the polarization \cite{FiSw2005,scullybook}
\begin{align}
 \bm{P}_{41} = & N \left( \bm{d}_{14} \varrho_{41} + \textrm{ c.c.} \right)\,,
\end{align}
with the definition of the susceptibility \cite{boydbook}
\begin{align}
 \bm{P}_{41} = & \varepsilon_0 \frac{E_{41}}{2} \left( \chi^{(1)} + \frac{3}{4} E_{41}^2 \chi^{(3)} \right)
 \hat{\bm{e}}_{41} e^{-i \omega_{41} t} + \textrm{ c.c.}\,,
\end{align}
we find
\begin{align}
 \chi^{(1)}(\omega_{41}) = & \frac{3}{8 \pi^2} \lambda_{41}^3 N \gamma_{41} \left[R_1^{(1)} \right]_{13}\,,\\
 \frac{3}{4} E_{41}^2 \chi^{(3)}(\omega_{41}) = & \frac{3}{8 \pi^2} \lambda_{41}^3 N \gamma_{41} \Omega_{41}^2 \left[R_3^{(1)} \right]_{13}\,,
\end{align}
with $\varepsilon_0$ being the permittivity of free space, $\lambda_{41}$ the wave length of the probe field transition, and $N$ the density of atoms in the gas.

We remark that $\chi^{(3)}(\omega_{41}) = \chi^{(3)}(\omega = \omega_{41} - \omega_{41} + \omega_{41})$ is the lowest order nonlinear contribution at the probe field frequency. It leads to an intensity dependent refractive index that also depends on $\omega_{41}$ and can be different for each respective frequency of the probe pulse spectrum. This is not the case for other contributions to $\chi^{(3)}$. For example, $[R_0^{(0)}]_{13}$ oscillates at the frequency $\omega = \omega_{41} - \Delta$ and leads to a contribution $\chi^{(3)}(\omega = \omega_{31} - \omega_{32} + \omega_{42})$ (four-wave mixing). Here, the resulting frequency is independent of $\omega_{41}$. Nevertheless, in principle those processes can influence the
result for the linear and third-order susceptibility at certain probe field frequencies. For example, light can be scattered into the probe field mode
via different processes. Whether this or similar contributions change the probe pulse depends on the pulse's frequency width compared to the multiphoton detuning $\Delta$ and more general also on the propagation direction of the probe field relative to the control fields. A definite answer to this question requires an analysis of the full pulse propagation dynamics through the medium which is beyond the scope of this work.

\subsection{\label{dop}Doppler Broadening}
A typical experimental setup to investigate the coherence properties of a laser driven atomic gas would be a gas cell with a dilute alkali-atom vapor. For a dilute atomic gas theoretical predictions for the linear and nonlinear susceptibility can be made on the basis of a single atom analysis. This greatly facilitates the theoretical analysis. However, in a dilute gas at room temperature or above the atoms move at velocities where the frequency shift due to Doppler effect cannot be neglected compared to the natural line width given by the radiative decay rate $\gamma$. To calculate the Doppler effect for a single field, we assume a Maxwell-Boltzmann velocity distribution in laser propagation direction with a most probable velocity given by~\cite{demtrlaser}
\begin{align}
 v_m = & \sqrt{\frac{2 k_B T}{m}}
\end{align}
with $k_B$ the Boltzmann constant, $T$ the temperature, and $m$ the mass of the atom. The non-relativistic Doppler frequency shift is given by
\begin{align}
 \omega_{\rm eff} = & \omega \left(1 - \frac{v}{c}\right)\,,
\end{align}
where $\omega_{\rm eff}$ is the shifted frequency seen by the moving atom, $\omega$ is the lab frame laser frequency, $v$ is the velocity of the atom in laser propagation direction, and $c$ is the speed of light. The Doppler shift effectively leads to an additional detuning $\Delta_\textrm{Dop}$ with a Gaussian distribution~\cite{demtrlaser} 
\begin{align}
 \label{distr}
 f(\Delta_\textrm{Dop}) \: d\Delta_\textrm{Dop} = & \frac{1}{\sqrt{\pi} k v_m}\, e^{-\left(\frac{\Delta_\textrm{Dop}}{k v_m}\right)^2} \: d\Delta_\textrm{Dop}\,,
\end{align}
where $k$ is the wave number. The corresponding line width (FWHM) is then given by
\begin{align}
\label{linewidth}
 \delta\omega = & k \sqrt{\textrm{ln}(2) \frac{8 k_B T}{m}}\,.
\end{align}
To actually calculate the linear and nonlinear susceptibility for a Doppler broadened medium, for each propagation direction, we have to  add $\Delta_\textrm{Dop}$ to the detuning of the fields propagating in this direction and then average the resulting susceptibility over the velocity distribution Eq.~(\ref{distr}).

\subsection{\label{buf}Buffer Gas and Pressure Broadening}
Introducing a buffer gas to the gas cell leads to more frequent collisions between the atoms. This has two main consequences. First of all it causes pressure broadening. For moderate densities, a collision between two atoms disturbs the level energies for a short time which results in the loss of phase coherence. In a simple approach this can be modeled by an additional decay rate $\gamma_c$ for the coherences. This collisional decay rate consists of a contribution due to the studied gas itself and a contribution due to the buffer gas. Both depend linearly on the respective density  \cite{boydbook},
\begin{align}
 \label{gamc}
 \gamma_c = & C_s N_s + C_b N_b\,,
\end{align}
with gas specific constants $C_s$ and $C_b$.

A second major effect of a buffer gas is closely connected to Doppler broadening. Due to the higher density the mean free path of a single atom moving in the gas is reduced. If it is reduced below the transition wavelength an averaging over different velocities during a single emission or absorption process can effectively re-narrow a Doppler broadened line. This phenomenon is known as Dicke narrowing \cite{dicke}.

\begin{figure}[t]
\includegraphics[width=8cm]{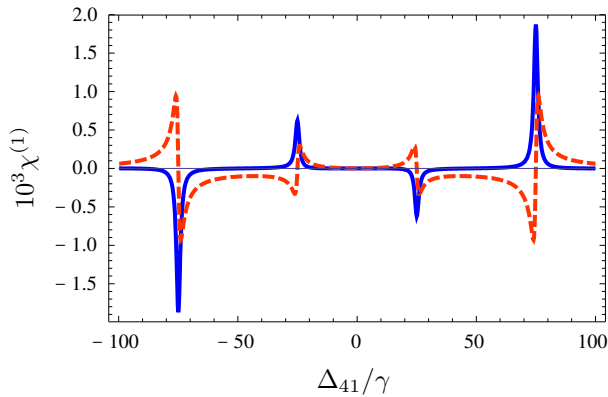}
\caption{\label{fig-splitting}(Color online) Real part (solid blue line) and imaginary part (dashed red line) of the linear susceptibility of the probe field. Due to strong control fields $\Omega_{42} = 100 \gamma$ and $\Omega_{31} = 50 \gamma$ the probe field resonance is split into four different resonances. Further, $\Omega_{32} = \Delta_{31} = \Delta_{32} = \Delta_{42} = 0$, and all spontaneous decay rates $\gamma_{jk}$ have been set to $\gamma$. The susceptibility is plotted in units of $3/8\pi^2 \lambda_{41}^3 N$.}
\end{figure}

\section{\label{res}Results}
In principle, Eqs.~(\ref{hie}) can be used to calculate analytical results for the desired $\chi^{(1)}$ and $\chi^{(3)}$. But in our situation of interest where all four electromagnetic fields, possibly all with different detuning, interact with the atom, these are usually to lengthy to give any physical insight. Therefore, we proceed with a numerical study of the linear and nonlinear susceptibility. 

\subsection{\label{wou}Without Doppler Broadening}
Here, our primary goal is to find a set of parameters where the intensity dependent refractive index is large enough to cause an appreciable amount of nonlinear selfphase modulation while the attenuation of a light pulse due to absorption is small. To achieve a high non-linear index of refraction with low linear and non-linear loss all in the same spectral region is challenging because resonances that enhance the nonlinear response typically  come with strong absorption. Still, we find such a suitable parameter set by manipulating the linear and nonlinear susceptibility of the probe field as described next.

We first split the unperturbed resonance of the probe field transition by a strong coupling field $\Omega_{42}$ and again about half as much by the second coupling field $\Omega_{31}$. This gives rise to four resonance structures in the linear response, see Fig.~\ref{fig-splitting}.

\begin{figure*}[t]
\includegraphics[width=14cm]{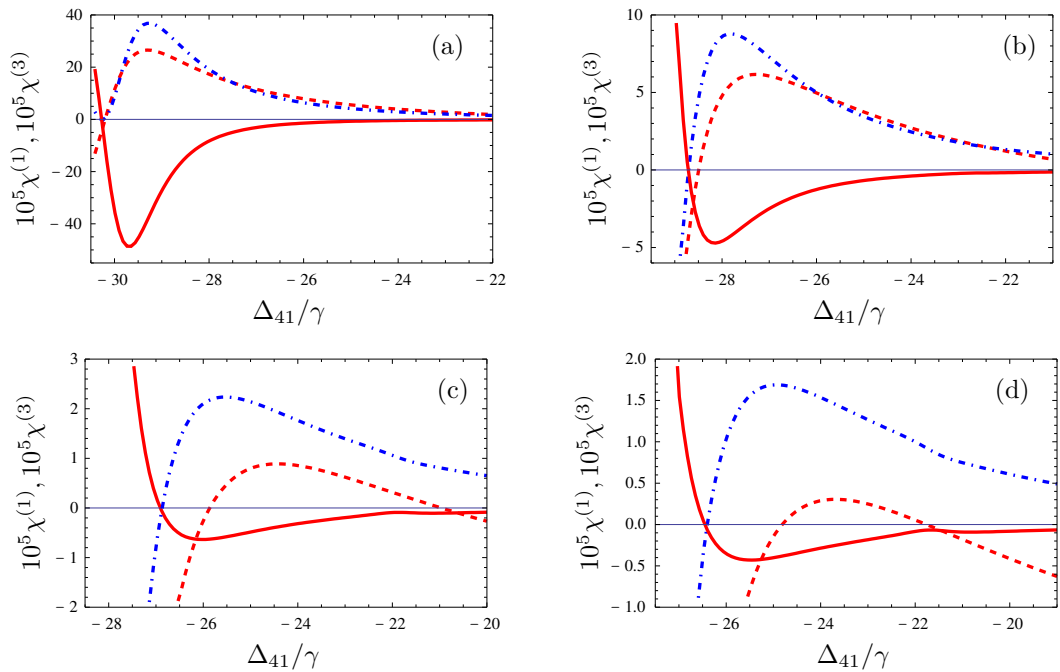}
\caption{\label{fig-res}(Color online) Real part (dash-dotted blue line) and imaginary part (solid red line) of the nonlinear susceptibility together with the imaginary part of the linear susceptibility (dashed red line). All figures show the resonance around $\Delta_{41} = -25 \gamma$. The susceptibility is plotted in units of $3/8\pi^2 \lambda_{41}^3 N$ and for comparability $\chi^{(3)}$ has been scaled with $3/4 E_{41}^2$. The parameters are $\Delta_{32} = \Delta_{42} = 0$, $\Omega_{31} = 50 \gamma$, $\Omega_{32} = 34 \gamma$, and $\Omega_{42} = 100 \gamma$. The probe field strength is assumed to be one tenth of the weakest control field in all cases. 
The detuning $\Delta_{31}$ is chosen as (a) $\Delta_{31} = 0$, (b) $\Delta_{31} = 0.7 \gamma$, (c) $\Delta_{31} = 1.5 \gamma$, and (d) $\Delta_{31} = 1.7 \gamma$. Note the different axis scales in the four subpanels.}
\end{figure*}
In this figure, the linear absorption of the resonance at $\Delta_{41} \approx -25 \gamma$ can be lowered by a small detuning $\Delta_{31}$, which modifies the dressed state populations. Finally, optimizing the result with the third coupling $\Omega_{32}$, we can tune one half of the resonance to a small linear and nonlinear absorption while still maintaining a substantial nonlinear real part. In Fig.~\ref{fig-res} it is shown how gradually introducing a detuning $\Delta_{31}$ influences the linear absorption, the nonlinear gain, and the real part of the nonlinear susceptibility. It decreases the linear absorption and the nonlinear gain faster than the real part and thereby improves their ratio. 
Interestingly, the imaginary parts of the linear and the nonlinear parts
of the susceptibility can have opposite signs in this spectral region .
The linear response induces absorption, while the nonlinear response
leads to gain. Absorption could in this spectral region therefore be reduced even further by a partial cancelling of linear absorption and nonlinear gain. However, these results are preliminary in the sense, that no effects due to Doppler and pressure broadening have been included yet.

\subsection{\label{wth}Including Doppler Broadening}
Using our considerations from Secs. \ref{dop} and \ref{buf} we now want to calculate the linear and nonlinear susceptibility in a Doppler broadened atomic gas. As a realistic example we want to assume a Sodium vapor with a density of $N = 1.0 \times 10^{20}\, \textrm{m}^{-3}$. To reach a vapor pressure that corresponds to this density the gas cell must be heated to a temperature of $T = 547.6$~K \cite{sodiumdata}. At this temperature the Doppler linewidth is $\delta\omega = 2 \pi \times 1.78$~GHz which is very broad compared to the natural linewidth of the Sodium $\textrm{D}_1$ transition of $\gamma = 2 \pi \times 9.76$~MHz. In a pure Sodium vapor the spectral features we found in Sec.~\ref{wou} would be averaged out by the Doppler effect. But if we introduce a buffer gas strong pressure broadening can preserve them. 
For Argon and Sodium, the gas parameters in Eq.~(\ref{gamc}) are given by $C_s = 1.50 \times 10^{-13}\, \textrm{m}^3 \,\textrm{s}^{-1}$ and $C_b = 2.53 \times 10^{-15}\, \textrm{m}^3\, \textrm{s}^{-1}$~\cite{boydbook}. 
We want to assume a collision-induced coherence loss rate of $\gamma_c=1.0$~GHz which corresponds to a buffer gas density of $N_b=3.95 \times 10^{23} \textrm{m}^{-3}$. At such a density the mean free path is of order $\Lambda = 10^{-5}$~m. This is much larger than the transition wavelength $\lambda=589.2 \times 10^{-9}$~m such that the limit of Dicke narrowing is not reached.
\begin{figure*}[t]
\includegraphics[width=14.0cm]{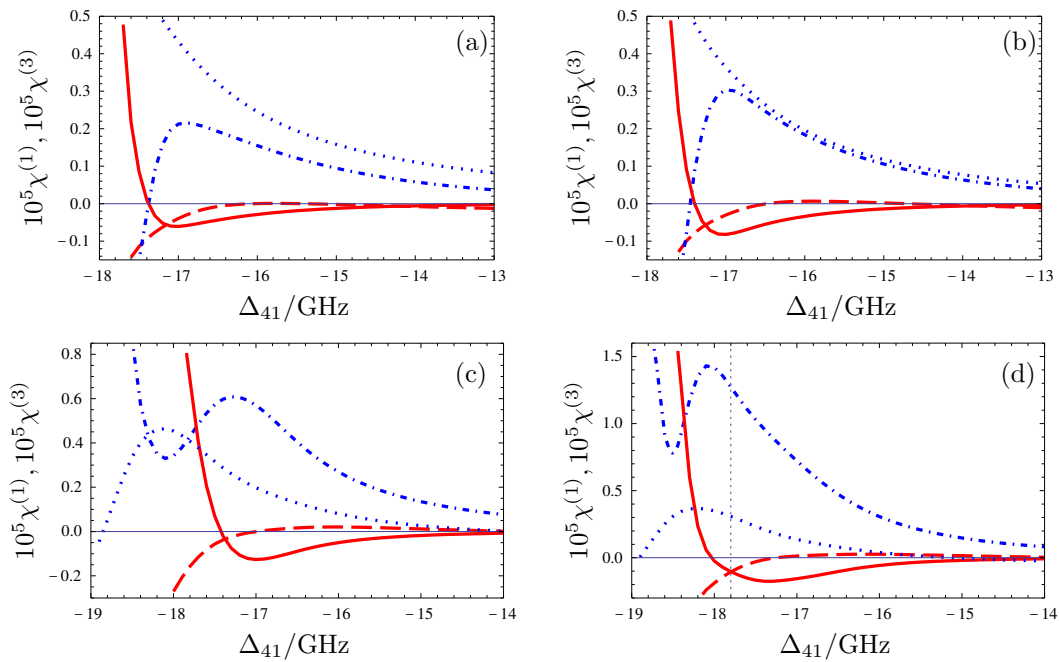}
\caption{\label{fig-dop}(Color online) Real part (dash-dotted blue line) and imaginary part (solid red line) of the nonlinear susceptibility together with the real part (blue dotted line) and the imaginary part of the linear susceptibility (dashed red line) at the resonance around $\Delta_{41} = -15.0$ GHz. The control fields have Rabi frequencies $\Omega_{42} = 60$~GHz, $\Omega_{31} = 30$~GHz, $\Omega_{32} = 25$~GHz,
and the detunings are $\Delta_{31}=1.6$~GHz, $\Delta_{32}=\Delta_{42}=0$.
The medium parameters described in the main text correspond to Sodium as the active medium with Argon as a buffer gas.  The four different plots show Doppler averaged results with a Doppler linewidth of (a)~below the natural linewidth, (b)~$50\%$, (c)~$90\%$, and (d)~$100\%$ of the full Doppler linewidth of $\delta\omega = 2 \pi \times 1.78$~GHz.}
\end{figure*}

We now try to recover results similar to the unbroadened case shown in 
Fig.~\ref{fig-res}.
Because of the strong broadening we have to apply correspondingly stronger control fields. For $\Omega_{42} = 60.0$~GHz and $\Omega_{31} = 30.0$~GHz, we find the resonance studied in the unbroadened case at around $\Delta_{41} = -15.0$~GHz. The third control field is set to $\Omega_{32} = 25.0$~GHz and the detuning to $\Delta_{31} = 1.6$~GHz. For the Doppler averaging we have assumed all fields to be co-propagating. 
The different subpanels in Fig.~\ref{fig-dop} correspond to different Doppler linewidths, and thus via Eq.~(\ref{linewidth}) to different temperatures.
In Fig.~\ref{fig-dop}(a), the Doppler linewidth is chosen below the natural 
linewidth of the probe transition, and as expected we finds results that are similar in shape to the unbroadened case (see Fig.~\ref{fig-res}(d)). Differences are mainly due to pressure broadening. Gradually increasing the Doppler linewidth up to the full Doppler width expected for the gas parameters discussed above in subfigure (d), we find that while the shapes of the different curves change, our main result of high nonlinear index of refraction with small linear and non-linear absorption persists with Doppler broadening. Also in the broadened case, a partial cancelling of linear absorption and nonlinear gain could be possible.
Note that since the averaging process affects not only the probe field detuning but all four detunings at the same time the results cannot be explained in terms of a simple  smoothing of the curves without Doppler effect. 

We also considered different laser geometries, such as control fields propagating perpendicular to the probe field, or one or two control field propagating in opposite directions, and found the co-propagating case to be the most advantageous one. This is similar to the case of Doppler broadening in typical electromagnetically induced transparency setups where co-propagating lasers typically are preferable.

We finally use our results at probe field detuning $\Delta = -17.8$ GHz to calculate the required optical length for a nonlinear selfphase modulation of $\pi$.
This probe field frequency is indicated by the vertical blue dotted line in Fig.~\ref{fig-dop}(d). The nonlinear selfphase modulation is given by~\cite{boydbook}
\begin{align}
 \Delta\Phi_\textrm{Nl} = & n_2\, I\, k \,L\,,
\end{align}
with $n_2$ the intensity dependent refractive index, $I$ the probe field intensity, $k$ the wavevector, and $L$ the optical length. We assume a probe field strength one tenth of the smallest control field and find
\begin{align}
 L_\pi = & 2.9 \textrm{ cm .}
\end{align}
From Fig.~\ref{fig-dop} (d) we see that the magnitude of the imaginary parts of the linear and nonlinear susceptibility are more than one order of magnitude smaller. 
Therefore, the equivalent characteristic length scale is more than one order of magnitude larger. Furthermore, both parts give rise to small gain rather than absorption.

Thus, our results show, that in a certain spectral region a nonlinear selfphase modulation of $\pi$ can be achieved on a realistic laboratory lengthscale. Since the real part of both the linear and the nonlinear susceptibility display approximately linear dispersion in the spectral region of interest,  pulse shape distortions can be expected to be small. Interestingly, the real part of the linear susceptibility has a negative slope in the considered frequency region, in contrast to a positive slope typically found in an electromagnetically induced transparency window.

\section{\label{con}Conclusion}
We have studied nonlinear effects in pulse propagation through a laser-driven medium where the applied fields form a closed interaction loop. Such loop systems in general only allow for a time-independent treatment at a single probe field frequency, where the so-called multiphoton resonance condition is fulfilled. As a probe field pulse has a finite frequency width, this condition which allows for a straightforward theoretical treatment could not be applied. Instead, we treated the time-dependent problem by turning it into a hierarchy of equations that describe the various physical processes occurring in the medium. We have included Doppler and pressure broadening as well as a buffer gas in our analysis and have used realistic parameters for a medium consisting of Sodium vapor. We could show that the studied system can exhibit a high non-linear refractive index with small absorption or gain over a spectral range of several natural line widths. For the chosen parameters, both the linear and the non-linear susceptibilities show near-linear dispersion such that pulse shape distortions are minimized, and the slope of the linear dispersion is negative.
A non-linear selfphase modulation of $\pi$ is obtained after $2.9$ cm propagation through the medium.

\appendix*
\section{Coefficients Matrix}
\label{coe}
The explicit form of the coefficient matrix $M$ and the inhomogeneous part $\Sigma$ can be derived from Eq.~(\ref{eom}). Here, we list all nonzero elements $M_{j,k}$ and $\Sigma_j$, which are given by

{\allowdisplaybreaks
\begin{align*}
M_{1,1} & = M_{1,6} = M_{6,6} = \frac{1}{2} M_{11,11}\\
& = \Sigma_1 = \Sigma_6\\
& = - \gamma_r \,,\\
M_{1,3}^* & = M_{1,9} = M_{2,10} = M_{3,4}\\
& = M_{4,12} = M_{5,7}^* = M_{9,11} = M_{13,15}^*\\
& = \frac{i}{2} \Omega_{31} \,,\\
M_{2,3}^* & = M_{5,9} = M_{6,7}^* = M_{6,10}\\
& = M_{7,11} = M_{8,12} = M_{10,11}^* = M_{14,15}^*\\
& = \frac{i}{2} \Omega_{32} \,,\\
M_{1,4}^* & = M_{1,13} = M_{2,14} = M_{3,15}\\
& = \frac{1}{2} M_{4,1}^* = M_{4,6}^* = M_{4,11}^* = M_{5,8}^*\\
& = M_{9,12}^* = \frac{1}{2} M_{13,1} = M_{13,6} = M_{13,11}\\
& = \Sigma_4^* = \Sigma_{13}\\
& = \frac{i}{2} \Omega_{41} e^{-i(\Delta t - \phi)} \,,\\
M_{2,4}^* & = M_{5,13} = M_{6,8}^* = M_{6,14}\\
& = M_{7,15} = M_{8,1}^* = \frac{1}{2} M_{8,6}^* = M_{8,11}^*\\
& = M_{10,12}^* = M_{14,1} = \frac{1}{2} M_{14,6} = M_{14,11}\\
& = \Sigma_8^* = \Sigma_{14}\\
& = \frac{i}{2} \Omega_{42} \,,\\
M_{3,3} & = M_{9,9}^*\\
& = - \gamma_r - i \Delta_{31} \,,\\
M_{4,4} & = M_{13,13}^*\\
& = - \gamma_r - i (\Delta_{31} + \Delta_{42} - \Delta_{32}) \,,\\
M_{7,7} & = M_{10,10}^*\\
& = - \gamma_r - i \Delta_{32} \,,\\
M_{7,8} & = M_{10,10}\\
& = - \gamma_r - i \Delta_{32} \,,\\
M_{12,12} & = M_{15,15}^*\\
& = -2 \gamma_r - i (\Delta_{42} - \Delta_{32}) \,,\\
M_{2,2} & = M_{5,5}^*\\
& = - i (\Delta_{31} - \Delta_{32}) \,,\\
M_{6,4} & = M_{11,4} = M_{6,13} = M_{11,13}\\
& = M_{1,8} = M_{11,8} = M_{1,14} = M_{11,14}\\
& = 0\,,
\end{align*}
}
where $M_{j,k} = M_{k,j}$ holds if not noted otherwise and  by $M_{j,k}^*$ we indicate the complex conjugate of $M_{j,k}$.


\end{document}